\documentclass{emulateapj}
\usepackage{apjfonts}
\lefthead{HAN \& HWANG} 
\righthead{MICROLENSING BINARIES}

\begin{document}
\title{The Importance of Binary Gravitational Microlensing Events
Through High-Magnification Channel}
\author{Cheongho Han and Kyu-Ha Hwang}
\affil{Department of Physics, 
Chungbuk National University, Cheongju 361-763, Republic of Korea;\\
cheongho,kyuha@astroph.chungbuk.ac.kr}


\begin{abstract}
We estimate the detection efficiency of binary gravitational lensing 
events through the channel of high-magnification events.  From this 
estimation, we find that binaries in the separations ranges of $0.1
\lesssim s\lesssim 10$, $0.2\lesssim s\lesssim 5$, and  $0.3\lesssim 
s\lesssim 3$ can be detected with $\sim 100\%$ efficiency for events 
with magnifications higher than  $A=100$, 50, and 10, respectively, 
where $s$ represents the projected separation between the lens 
components normalized by the Einstein radius.  We also find that 
the range of high efficiency covers nearly the whole mass-ratio range 
of stellar companions.  Due to the high efficiency in wide ranges 
of parameter space, we point out that majority of binary-lens events 
will be detected through the high-magnification channel in lensing 
surveys that focus on high-magnification events for efficient 
detections of microlensing planets.  In addition to the high 
efficiency, the simplicity of the efficiency estimation makes the 
sample of these binaries useful in the statistical studies of the 
distributions of binary companions as functions of mass ratio and 
separation.  We also discuss other importance of these events.
\end{abstract}



\section{Introduction}

If a star is gravitationally microlensed by a lens system 
composed of two masses, the resulting light curve can 
dramatically deviate from the smooth and symmetric one 
of a single-lens event.  This deviation is caused by 
the formation of caustics for binary-lens systems. The 
caustics represent the source positions at which the 
lensing magnification of a point source becomes infinite.  
The set of caustics forms one, two, or three close curves 
each of which is composed of concave curves that meet at 
points.  By analyzing the light curve of a binary-lens 
event, it is possible to obtain information about 
the lens system because the structure of the caustic 
system and the resulting light curve vary depending 
on the mass ratio and the projected separation between 
the components of binary lenses.

Since the pioneering work by \citet{chang80, chang84}, 
binary lensing has been a subject of intense theoretical 
studies.  \citet{schneider86} made a comprehensive study 
of binary lenses in order to learn about caustics in 
quasar macrolensing.  \citet{witt90} developed a simple 
algorithm for finding caustics of binary lenses.  \citet{witt95} 
studied lensing magnification inside caustic and found that 
the minimum magnification when the source is inside a caustic 
is greater than 3.  \citet{rhie97} found that the maximum 
number of images for multiple-lens systems.  With the 
beginning of microlensing surveys, theoretical studies 
became even more active.  \citet{gaudi97} pointed out 
that microlensing is an efficient method to detect close 
binaries.  \citet{distefano97} mentioned various channels 
of detecting binaries including repeating events.  
\citet{dominik99b} studied the lensing behavior in the 
extreme cases of binary separations and mass ratios.  
\citet{dominik99a} and \citet{albrow99b} mentioned possible 
degeneracies in modeling light curves of binary-lens events.  
\citet{han99} and \citet{han01} studied the astrometric 
behavior of binary-lens events.  \citet{bozza00,bozza01} 
derived analytic expressions for the location of caustics 
and studied the motion of images of microlensed stars.  
\citet{gaudi02a, gaudi02b} investigated the photometric 
and astrometric behaviors in the region very close to 
caustics.  \citet{graff02} devised a method to measure 
the mass of the binary-lens system from the analysis of 
light curves of caustic-crossing events.  In addition to 
the theoretical studies, binary-lens events were actually 
detected from various surveys \citep{udalski94, udalski98, 
alcock99, alard95, afonso00, albrow99a, albrow00, albrow01, 
an02, smith02, albrow02, abe03, kubas05, jaroszynski04, cassan04, 
jaroszynski05, jaroszynski06}.

With the active researches in both theoretical and 
observational fields, binary microlensing has developed 
into a useful tool to study stellar astrophysics.  The 
most active field of application is the stellar atmosphere 
for which microlensing is used to probe detailed structures 
on the surface of source stars by using the high resolution 
of caustic-crossing events \citep{albrow99a, albrow01, 
abe03}.

Microlensing can also be used to probe the distributions 
of binary companions of Galactic stars as functions of 
mass ratio and separation.  These binary distributions 
provide important observational constraints on theories 
of star formation.  Since microlensing is sensitive to 
low-mass companions that are difficult to be detected 
by other methods, it is in principle possible to make 
complete distribution down to the lower mass limit of 
binary companions.  Despite the importance, the progress 
of this application of binary lensing has been stagnant.  
There are two main reasons for this.  The first reason 
is caused by the difficulty in estimating the detection 
efficiency of binary-lens events.  In previous lensing 
surveys, most binary-lens events were discovered through 
the channel of caustic-crossing events, in which the caustic 
crossings were accidently discovered from the sudden rise 
of the source star flux.  Due to the haphazard nature of 
caustic crossings, it was difficult to estimate the detection 
efficiency that is essential for the statistical studies of 
binary companions.  The second reason is that microlensing 
is mainly sensitive to binaries over a narrow range of 
projected separations.  This limits especially the study 
of the distribution of binary separations.  

When the first microlensing surveys \citep{alcock93, 
aubourg93} were started, the prime scientific goal was  
constraining the Galactic dark matter in the form of massive 
compact halo objects.  However, the most important science 
of the current lensing surveys \citep{udalski03, bond02, 
beaulieu98, yoo04} is detections of extrasolar planets.  
With the change of the goal, the observational strategy 
of the surveys also changed.  These changes include the 
employment of early warning system and the operation of 
follow-up observations.  Another important change is the 
adoption of an observational strategy focusing on 
high-magnification events for which the probability of 
planet detection is high \citep{griest98}.

In this paper, we emphasize the importance of binary-lens 
events to be detected through the high-magnification channel, 
especially for the statistical studies of Galactic binaries.  
We demonstrate the high efficiency of the high-magnification 
channel to binaries in a wide range of parameter space
and the simplicity of the efficiency estimation.  We also 
discuss other importance of these events.

\section{Binary Caustics}

The number, size, and shape of caustics induced by binary 
lenses vary depending on the mass ratio, $q$, and the 
projected separation, $s$, between the lens components in 
units of the Einstein radius corresponding to the the total 
mass of the binary, $\theta_{\rm E}$.  It is customary to 
divide the topology of binary caustics into three categories 
of intermediate, close, and wide binaries.

Intermediate binaries refer to the case for which the separation 
between the lens components is in the range 
\citep{erdl93}
\begin{equation}
(1+q)^{1/4}(1+q^{1/3})^{-3/4} < s< (1+q)^{-1/2}
(1+q^{1/3})^{3/2}.
\label{eq1}
\end{equation}
In this case, there exists a single large caustic with six 
cusps.  Caustics induced by intermediate binaries are often 
referred as resonant caustics because they are produced when 
the binary separation is similar to the Einstein radius of 
the lens system.

Close binaries refer to the case for which the separation is 
smaller than those of intermediate binaries.  In this case, 
there are a single major caustic with four cusps and two 
small outlying caustics with three cusps.  The center of 
the major caustic is located approximately at the 
center-of-mass of the binary system.  The size of the 
central caustic becomes smaller as the separation becomes 
smaller.  In the limiting case of a close binary with 
$s\ll 1$, the lensing behavior is approximated as that of 
a single lens with a mass equivalent to the total mass of the 
binary and the location at the center of mass of the binary.

Wide binaries refer to the case for which the separation is 
larger than those of intermediate binaries.  In this case, 
there exist two four-cusp caustics each of which is associated 
with each lens component.  The size of each caustic becomes 
smaller as the separation increases.  In the limiting case 
of $s\gg 1$, the lensing behavior is approximated as that 
of two independent single lenses located at the positions 
of the individual lens components.

\section{Detection Efficiency}

In this section, we demonstrate that high-magnification
events provide an important channel to detect binary 
companions in wide ranges of binary parameters.  For this, 
we construct the distributions of efficiency as functions 
of the binary separation and the mass ratio.

The efficiency is estimated as follows.  For a binary 
with a given set of the projected separation $s$ and the 
mass ratio $q$, we produce many light curves of binary-lens 
events resulting from source trajectories with impact 
parameters and orientation angles with respect to the binary 
axis distributed in the ranges of $u_0\leq u_{0,{\rm th}}$ 
and $0\leq \alpha\leq 2\pi$, respectively.  Here $u_{0,{\rm th}}$ 
represents the threshold impact parameter to the source 
trajectory corresponding to a threshold magnification 
$A_{\rm th}$, which represents the minimum magnification 
required for intensive follow-up observations.  For each 
binary-lens light curve, we also produce a single-lens 
light curve that approximates the light curve of the binary 
event.  We then compare the two light curves and compute 
the fractional deviation $\epsilon=(A-A_{\rm s})/A_{\rm s}$, 
where $A$ and $A_{\rm s}$ represent the lensing magnifications 
of the binary and single-lens events, respectively. We then 
estimate the detection efficiency as the fraction of events 
with fractional deviations greater than a threshold value, 
$\epsilon_{\rm th}$, among the total number of tested events.  
To see the distribution of efficiency, we repeat the 
procedure for binaries with different values of $s$ and $q$.

Caustics induced by close or wide binaries can be very small
and thus perturbations induced by such caustics are 
vulnerable to the effect of finite source size. We therefore
consider the effect in our efficiency computation.  For this, 
we use the ray-shooting technique \citep{schneider86, kayser86, 
wambsganss97, dong06}.  In this technique, a large number of 
light rays are uniformly shot from the observer plane through 
the lens plane and then collected on the source plane and the 
magnification map of a region on the source plane is obtained 
by the ratio of the number densities of rays on the source plane 
to that on the lens plane.  With the constructed magnification 
map, the light curve of an event is obtained from a one-dimensional 
cut on the map.  We accelerate this process by restricting the 
region of ray-shooting only in the region around the caustic 
and use a simple semi-analytic approximation in other parts 
of the source plane \citep{pejcha09, gould08}.  In addition, 
we keep the information of the positions of the light rays 
arriving at the target in the buffer memory of a computer so 
that it can be readily used for fast computation of magnifications.
For the source radius normalized by the Einstein radius, we use 
$\rho_\star=0.003$ by adopting the value of a typical Galactic 
bulge event occurring on a bright main-sequence source star.

\begin{figure}[t]
\epsscale{1.2}
\plotone{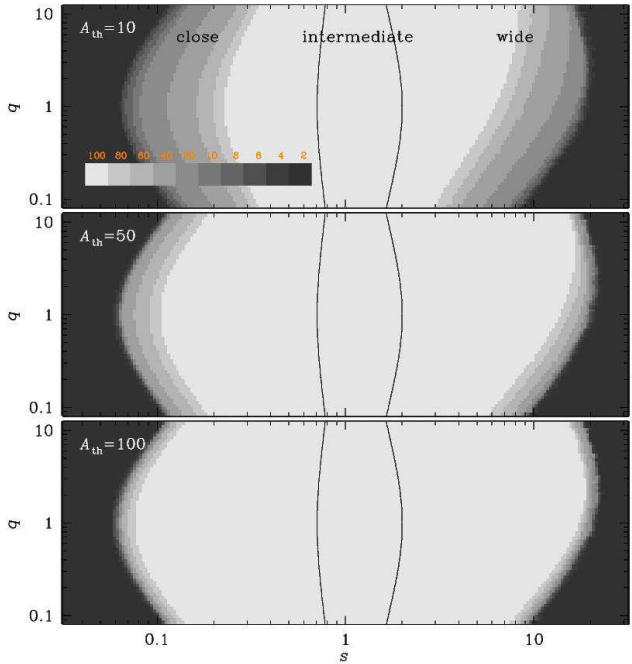}
\caption{\label{fig:one}
Distributions of the binary-detection efficiency as grey-scale 
maps in the parameter space of $(s,q)$.  The individual panels 
show the distributions for three different threshold magnifications 
of $A_{\rm th}=10$ (upper panel), 50 (middle panel), and 100 
(lower panel), respectively. The efficiency is given in percentage.
The solid curves in the middle of each diagram represent the 
boundaries between close and intermediate and 
wide and intermediate binaries.
}\end{figure}

Figure~\ref{fig:one} shows the constructed distributions of the 
efficiency as grey-scale maps in the parameter space of $(s,q)$ 
for three different threshold magnifications of $A_{\rm th}=10$, 
50, and 100.  We note that the cases with $q<1$ and $q>1.0$ 
represent that the source trajectory approaches closer to the 
heavier and lighter component of the binary, respectively.  
We set the threshold deviation as $\epsilon_{\rm th}=0.1$.

From the distributions, we find that the efficiency is high 
for wide ranges of separation and mass ratio.  For a 
threshold magnification of $A_{\rm th}=100$,  we find that 
the efficiency is $\sim 100\%$ for binaries with separations 
in the range of $0.1\lesssim s \lesssim 10$.  The region of 
high efficiency is substantial even for lower threshold 
magnifications.  We find that the ranges of 100\% efficiency 
are $0.2\lesssim s \lesssim 5$ and $0.3\lesssim s \lesssim 3$ 
for events with magnifications greater than $A_{\rm th}=50$ and 
10, respectively.   We also find that the range of high efficiency 
covers nearly  the whole mass-ratio range of stellar companions.

The high efficiency is due to the existence of central caustics 
induced by close or wide binaries.  The lensing behavior of a 
wide binary with $s\gg 1$ is well described by the Chang-Refsdal 
lensing.  In this regime, the width of the caustic is approximated 
as
\begin{equation}
\xi_{\rm c}\sim {4q \over s^2(1+q)}.
\label{eq2}
\end{equation}
Then, for a wide-separation binary with $s=10$ and $q=1.0$, 
the caustic width is $\xi_{\rm c} \sim 0.02$ as measured by
$\theta_{\rm E}$. This corresponds to the $\hat{\xi} \sim 0.03$ 
as measured by the Einstein radius corresponding to the mass of 
the binary component to which the source trajectory approaches 
more closely, $\hat{\theta}_{\rm E}$.  The perturbation extends 
outside the caustic.  Assuming that the region of detectable 
perturbation extends twice of the caustic size, it is found that 
perturbations can be detected for events with $A\gtrsim 30$.  
For close binaries, the caustic size of a binary with a separation 
$s$ is equivalent to the caustic size of a wide binary with a 
separation $s^{-1}$.  Therefore, the lower limit of the separation 
range roughly corresponds to the inverse of the upper 
limit.\footnote{Actually, 
there exists asymmetry between the efficiency distributions of 
the regions with $s<1$ and $s>1$. This asymmetry is caused by 
the difference of the single-lens approximations between close 
and wide binaries.  For an event produced by a close binary, 
the light curve of the event is approximated as that of a single 
lens with a mass corresponding to the {\it total} mass of the 
binary.
For an event produced by a wide binary, on the other hand, 
the light curve is approximated as the light curve produced 
by a single-lens event with a mass corresponding to the mass 
of the lens component that the source trajectory approaches 
closer and thus the Einstein radius is smaller by 
$\hat{\theta}_{\rm E}=\theta_{\rm E}/(1+q)^{1/2}$.
Then, although the size of the two caustics induced by a close 
binary and a wide binary with separations $s$ and $s^{-1}$ are 
equivalent as measured by $\theta_{\rm E}$, the caustic size 
measured by $\hat{\theta}_{\rm E}$ is bigger. As a result, 
the efficiency is higher for wide binaries compared to close 
binaries with corresponding inverse separations.
The asymmetry between the distributions in the regions of 
wide binaries with $q<1$ and $q>1$ is caused by the fact that 
heavier companions induce larger caustics.  This asymmetry 
does not occur for close binaries because the two masses are  
close and thus the source does not distinguish the two.}

\section{Implication}

The high efficiency of the high-magnification channel in 
detecting binary companions has several important scientific 
implications.  First, it implies that high-magnification events 
will provide a major channel of detecting binaries in lensing 
surveys that focus on high-magnification events for efficient 
detections of microlensing planets.  Due to the location of 
perturbations similar to those induced by planets, perturbations 
induced by binary companions will be densely monitored by follow-up 
observations.  In addition, the perturbations will be observed with 
high precision since they occur when the magnification is high.  
Being observed with high time resolution and photometric precision, 
it will be possible to accurately characterize the physical 
parameters of the individual binary systems.

Second, the sample of binaries to be detected through 
high-magnification channel will makes it possible to study the 
binary distributions due to the simplicity of efficiency estimation.  
Unlike the accidental detections of binary events in previous 
surveys, binaries probed by the high-magnification channel are 
detected under a simple criterion that the peak magnification 
is greater then a certain threshold.  As a result, it will be 
possible to statistically investigate the binary properties.

Another importance of the high-magnification channel is that 
it enables to detect planets in binary systems.  This is 
possible because both planet and binary companion produce 
perturbations in a common central region and thus the signatures 
of both companions can be simultaneously detected in the light 
curves of high-magnification events.  Two types of planets in 
binaries can be detected.  The first type is a planet orbiting 
one of a wide binary system \citep{lee08}.  The other type is 
a planet orbiting a close binary system \citep{han08}.  Planets 
of the former type were discovered by radial velocity surveys 
\citep{konacki05, eggenberger06}.  However, no firm detection 
of planets of the latter type has been reported because radial 
velocity surveys avoid short-term binaries as target stars.  
The chance to detect such planets by using the transit method 
is also very low because these planets tend to have wide orbits.  
Therefore, microlensing is an important method for the discoveries 
of planets with two simultaneously rising suns.

\begin{deluxetable*}{lccc}
\tablecaption{Optimal Binary Separation\label{table:one}}
\tablewidth{0pt}
\tablehead{
\multicolumn{1}{c}{event} &
\multicolumn{1}{c}{$s$} &
\multicolumn{1}{c}{$q$} &
\multicolumn{1}{c}{$u_0$} 
}
\startdata
OGLE-2009-BLG-020/MOA-2009-BLG-011 & 0.41      & 0.24      & 0.052       \\
MOA-2009-BLG-016                   & 0.21/5.98 & 0.18/0.25 & 0.019/0.017 \\
MOA-2009-BLG-137/OGLE-2009-BLG-092 & 2.91      & 0.57      & 0.096       \\
MOA-2009-BLG-273                   & 0.36      & 0.79      & 0.093       \\
MOA-2009-BLG-302                   & 0.51      & 0.22      & 0.066       \\
MOA-2009-BLG-406                   & 4.38      & 1.42      & 0.031       \\
MOA-2009-BLG-408                   & 0.16/9.94 & 0.33/0.95 & 0.003/0.002 
\enddata 
\tablecomments{
Binary-lens events detected through the high-magnification channel
in 2009 microlensing survey season along with the lensing parameters.
For the case for which the close/wide degeneracy is not clearly 
broken, two sets of parameters of the close/wide binary
solutions are listed.
}
\end{deluxetable*}

\section{Discussion}

To see the efficiency of high-magnification channel to binary 
events in actual lensing surveys, we search the data of microlensing 
surveys in 2009 season for binary events.  From this search, we 
find that there exist 7 binary-lens events that were detected 
through the high-magnification channel.  We list the events in 
Table~\ref{table:one} along with the lensing parameters.  The fact 
that these events comprise 70\% of all detected binary-lens events 
corroborates our claim that high-magnification events provide the 
major channel of detecting binary-lens events in current lensing 
surveys.  For all of these events, the perturbations were densely 
observed by follow-up observations, enabling to characterize the 
binary systems up to the well-known close/wide degeneracy.

The fact that central perturbations produced by a planet and a
binary companion occur near the peak of light curve brings up a 
question of the possibility to distinguish the two perturbations 
of different origins.  Although both binary and planetary systems 
can produce central caustics, the resulting perturbations induced 
by the two lens populations are different.  The basis of this 
difference lies in the difference in shape between the caustics 
of the two lens populations.  The shape of binary caustics varies 
such that the caustic is elongated along the binary axis when the 
separation is of order of the Einstein radius and it becomes 
more symmetric as the separation deviates from the Einstein radius. 
For a low-mass companion such as a planet, the companion should 
be located close to the Einstein radius to produce noticeable 
perturbations.  As a result, the central caustic induced by a 
planet has an elongated shape.  On the other hand, the central 
caustic induced by a close/wide binary companion has a symmetric 
diamond shape.  Therefore, detailed modeling of observed light 
curves enables to discriminate between the two interpretations.

In many cases, the two types of perturbations can be immediately
distinguished from characteristic features.  Several such diagnostic 
features were already proposed.  \citet{hangaudi08} proposed a 
diagnostic based on the difference in the shape of the intrapeak 
region of double-peaked high-magnification events.  They found 
that the shape of the intrapeak region is smooth and concave for 
binary lensing, while it tends to be either boxy or convex for 
planetary lensing due to the existence of the small, weak cusp 
of the planetary central caustic located between the two stronger 
cusps.  \citet{han09} proposed another diagnostic that can be 
applicable to perturbations affected by severe finite-source 
effect. He found that the feature induced by a binary companion 
forms a complete annulus, while the feature induced by a planet 
appears as several arc segments.  Then, the absence of a 
well-developed dip in the residual from the single-lensing light 
curve at either of the moments of the caustic center's entrance 
into and exit from the source star surface indicates that the 
perturbation is produced by a planetary companion.  \citet{hanetal09} 
found that a short time-scale caustic-crossing feature occurring 
at a moderate magnification with an additional secondary perturbation 
is a typical feature of binary-lens events and thus can be used as 
a diagnostic to discriminate between the binary and planetary 
interpretations.

\section{Conclusion}

In this paper, we emphasized the importance of binary-lens events 
to be discovered through the channel of high-magnification events.
Due to the high efficiency in wide ranges of parameter space, we 
pointed out that majority of binary-lens events will be detected 
through the high-magnification channel in current planetary 
microlensing surveys, In addition to the high efficiency, the 
simplicity of the efficiency estimation makes the sample of these 
binaries useful in the statistical studies of the distributions 
of binary companions as functions of mass ratio and separation.
We also discussed other importance of these events.

\acknowledgments 
This work is supported by Creative Research Initiative program
(2009-0081561) of National research Foundation of Korea.  
We would like to thank A.\ Gould and B.\ S.\ Gaudi for providing
helpful comments.

\end{document}